\documentclass[12pt]{article}
\usepackage{amsmath}
\usepackage{amsfonts}
\usepackage{amssymb}
\usepackage{latexsym}
\usepackage{graphicx}
\usepackage[english]{babel}
\input epsf.sty

\begin{document}

\title{{\normalsize \hskip4.2in {CAS-KITPC/ITP-017}} \\
{Inflationary NonGaussianity from Thermal Fluctuations}}
 \vspace{3mm}
\author{{Bin Chen$^{1}$, Yi Wang$^{2,3}$, Wei Xue$^{1}$}\\
{\small $^{1}$ Department of Physics, Peking University, Beijing 100871, P.R.China}\\
{\small $^{2}$ Institute of Theoretical Physics, Academia Sinica,
Beijing 100080, P.R.China}\\
{\small $^{3}$  The Interdisciplinary Center for Theoretical Study, University of }\\
{\small Science and Technology of China (USTC), Hefei, Anhui 230027,
P.R.China}}
\date{}
\maketitle

\begin{abstract}
We calculate the contribution of the fluctuations with the thermal
origin to the inflationary nonGaussianity. We find that even a small
component of radiation can lead to a large nonGaussianity. We show
that this thermal nonGaussianity always has positive $f_{\rm NL}$.
We illustrate our result in the chain inflation model and the very
weakly dissipative warm inflation model. We show that $f_{NL}\sim
{\cal O}(1)$ is general in such models. If we allow modified
equation of state, or some decoupling effects, the large thermal
nonGaussianity of order $f_{\rm NL}>5$ or even $f_{\rm NL}\sim 100$
can be produced. We also show that the power spectrum of chain
inflation should have a thermal origin. In the Appendix A, we made a
clarification on the different conventions used in the literature
related to the calculation of $f_{\rm NL}$.
\end{abstract}

\newpage

\section{Introduction}
Inflation has been remarkably successful in solving the problems
in the standard hot big bang cosmology
\cite{Guth81,Linde82,Steinhardt82,Starobinsky-inf}. Furthermore
inflation shows us that the fluctuations of quantum origin were
generated and frozen to seed the wrinkles in the cosmic microwave
background (CMB) \cite{CMBobserve,WMAP} and today's large scale
structure
\cite{Mukhanov81,Guth82,Hawking82,Starobinsky82,Bardeen83}. Its
prediction of a scale invariant spectrum which has been confirmed
in the experiments in the past decade is remarkable and has been
taken to be a great success of the theory.

Since the idea of inflation was proposed, there have been a large
number of inflation models. It has become one of the key problems in
cosmology to extract more information from experiments in order to
distinguish these inflation models. The key quantities from
experiments include the power spectrum of scalar and tensor
perturbations, the scalar spectral index and its running, and the
nonGaussianity.

The amount of nonGaussianity is often estimated using the quantity
$f_{\rm NL}$, which can be written as\footnote{Note that for the
definition of $f_{\rm NL}$, there is a sign difference between the
notation of the WMAP group \cite{Spergel:2006hy} and Maldacena's
calculation \cite{NG}. We use the same notation as that of the WMAP
group here. The difference is discussed in detail in the Appendix
A.}
\begin{equation}
\zeta=\zeta_g+\frac{3}{5}f_{\rm NL}\left({\zeta}_g^2-\langle
{\zeta}_g^2\rangle\right) ~,
\end{equation}
where the subscript $g$ denotes the Gaussian part of $\zeta$.

It has been shown that in the simplest single field slow roll
inflation models, the nonGaussianity estimator $f_{\rm NL}\sim {\cal
O}(\epsilon, \eta')$ \cite{NG, NG2}, where $\epsilon$ and $\eta$ are
the slow roll parameters. Such a small nonGaussianity is not only
much smaller than the current observational bound $|f_{\rm
NL}|<100$, but also well below the sensitivity of the Planck
satellite, $f_{\rm NL}\sim 5$. However, the recent study of the
inflation models with significantly nonlinear dynamics shows that
the nonGaussianity in them could be large. For example, in the DBI
\cite{DBI}, K-inflation \cite{K}, and ghost inflation \cite{Ghost}
models, $f_{\rm NL}$ can reach ${\cal O}(1)$ or larger than ${\cal
O}(1)$ in some parameter regions. Such a large nonGaussianity is
hoped to be observed in the future experiments and shed light on the
physics behind these models.

Recently, Yadav and Wandelt has claimed that from the WMAP 3-year
data, $f_{\rm NL}$ is detected at above 99.5\% confidence level
\cite{Yadav:2007yy}. They show that at 95\% confidence level, the
local shape $f_{\rm NL}$ is in the region
\begin{equation}
  26.91<f_{\rm NL}< 146.71~.
\end{equation}
If this result is confirmed by the WMAP 5-year data and Planck, a
great number of inflation models (without extra mechanisms) will be
ruled out, and there is also hope to measure the shape of $f_{\rm
NL}$, and the tri-spectrum $\tau_{\rm NL}$ at Planck.

In this paper, we propose another mechanism to produce potentially
large nonGaussianity. Instead of producing nonGaussianity from the
nonlinear evolution of the inflaton, in our mechanism, the large
nonGaussianity stems from the correlation in the initial conditions.
We will show that if the initial condition of the perturbations is
prepared in part by thermal fluctuations, there can be strong
3-point correlation, inducing large nonGaussianity.

In many inflation models, the radiation component only takes a very
small part in the energy density. But since the nonGaussianity from
the coherent motion of inflaton is highly suppressed,
the thermal nonGaussianity can play a significant part in $f_{\rm
NL}$, and in some parameter regions it provides the dominant
contribution. This may open a window for us to study the thermal
fluctuations in the models.

Indeed, the thermal effects are significantly important in some
inflation models. One example is the chain inflation. Based on the
the rapid tunnelling mechanism for the meta-stable vacua in the
string landscape\cite{HenryTye:2006tg, Podolsky:2007vg}, Freese,
Spolyar and Liu \cite{Freese:2004vs} proposed the so-called chain
inflation model in which the meta-stable vacua during inflation
tunnel very rapidly. The density perturbation in chain inflation is
calculated by Feldstein and Tweedie in \cite{Feldstein:2006hm} and
a simplified version of the chain inflation was proposed by Huang in
\cite{Huang:2007ek}.

In the chain inflation models, the average life time for a
meta-stable vacuum is much smaller than the Hubble time, so that the
vacuum decay via bubble nucleation takes place very rapidly, and
there can be many bubbles nucleated within one inflationary horizon.
These bubbles eventually collide and the energy stored in the bubble
wall decays into the radiation. This is very different from the slow
roll inflation models in which the decrease of the inflaton energy
density is wasted by the cosmic fraction, with very little radiation
being left.

Another example with large thermal effect is the warm inflation by
Berera and Fang \cite{Berera:1995wh}. In the warm inflation model,
a fraction of inflaton energy decays into radiation continuously
during inflation. The decay from inflaton to radiation can be
achieved by a interaction term in the inflaton's Lagrangian. It is
shown in warm inflation that due to the continuously creation of
radiation, the temperature during inflation can be nearly constant
\cite{Berera:1995wh}, so it provides a playground for
investigating the thermal effects. Previously, the nonGaussianity
of the warm inflation model is studied in \cite{Gupta:2002kn}. But
in \cite{Gupta:2002kn}, the authors considered only the
nonGaussianity of the inflaton field with Gaussian noise source,
and the nonGaussianity of the thermal fluctuation has not been
investigated. In this paper, we only consider the warm inflation
in the very weakly dissipative regime, in which the existence of
the thermal bath would not spoil the quantum vacuum.

In such inflation models with thermal radiation, it can be shown
that the nonGaussianity estimator $f_{\rm NL}$ is no longer
suppressed by the slow roll parameters. Even when the radiation
component is so tiny that it does not qualitatively change the
inflationary background, considerable nonGaussianity $f_{\rm NL}\sim
{\cal O}(1)$ can be produced.

Furthermore, we suppose in some cases, a new scale related to the
acoustic horizon, or some decoupling scales enters the calculation.
In this case, very large nonGaussianity of the order $f_{\rm NL}>5$
or even $f_{\rm NL}\sim 100$ can be produced without fine-tuning.

This paper is organized as follows. In Section 2, we develop the
general method to calculate the nonGaussianity of the thermal
origin. We calculate the 2-point and 3-point correlation functions
of thermal fluctuations. Based on these, we derive the power
spectrum and the nonGaussianity estimator $f_{\rm NL}$. In Section
3, we calculate the amount of nonGaussianity explicitly in
 the chain inflation model and the thermal
inflation model. We conclude in Section 4.

\section{The thermal correlation functions and the nonGaussianity}

In this section, we calculate the correlation functions, the power
spectrum and the nonGaussianity of thermal fluctuations. We also
give a simple estimate of the nonGaussianity by calculating the
backreaction.

We suppose the energy density takes the form
\begin{equation}\label{rho}
  \rho\equiv \rho_0+\rho_r=\rho_0+AT^m~,
\end{equation}
where $\rho_0$ is the energy density without thermal origin, for
example, the effective vacuum energy provided by the inflaton
potential. And $\rho_r$ is the energy density for the radiation, and
$A$ is a constant with dimension $[{\rm mass}]^{4-m}$. Note that
$m=4$ for usual radiation. While for generality, phenologically, we
still keep $m$ here.

The correlation functions in thermal equilibrium can be calculated
from the partition function of the system
\begin{equation}
  Z=\sum_r e^{-\beta E_r}~,
\end{equation}
where $\beta=T^{-1}$.

Let $U\equiv \rho V$ represents the total energy inside a volume
$V$. Then the average energy of the system is given by
\begin{equation}
  \langle U\rangle =  -\frac{d \log Z}{d\beta}~,
\end{equation}

The 2-point correlation function for the fluctuations $\delta
\rho\equiv \rho-\langle\rho\rangle$ is given by
\begin{equation}\label{rho2}
  \langle\delta\rho^2\rangle=\frac{\langle\delta
  U^2\rangle}{V^{2}}=\frac{1}{V^2}\frac{d^2 \log Z}{d\beta^2}=-\frac{1}{V^{2}}\frac{d\langle
  U\rangle}{d\beta}=\frac{mAT^{m+1}}{V}~,
\end{equation}
where in the final equality we have neglected ``$\langle\rangle$''
because the difference is next to the leading order.

Similarly, the 3-point correlation function can be expressed as
\begin{equation}\label{rho3}
  \langle\delta\rho^3\rangle=\frac{\langle\delta
  U^3\rangle}{V^{3}}=-\frac{1}{V^3}\frac{d^3\log
  Z}{d\beta^3}=\frac{1}{V^3}\frac{d^2\langle U\rangle}{d\beta^2}=\frac{m(m+1)AT^{m+2}}{V^2}~.
\end{equation}

Now let us apply the above calculation to inflation. First, we
calculate the equation of state $w_r$ for general radiation
$\rho_r=AT^m$. For simplicity, we only consider the case that $w_r$
is a positive constant. To do this, we temporarily consider
radiation without source. In the expanding background, consider a
comoving volume $V_c$ which is in the thermal equilibrium. The
conserved radiation entropy within this volume is given by
\begin{equation}\label{t1}
  S=\frac{\rho_r + p_r}{T}V_c~.
\end{equation}
The radiation energy density and pressure changes with respect to
the scale factor $a$ as
\begin{equation}\label{t2}
  p_r\sim \rho_r\sim a^{-3(w_r+1)}~.
\end{equation}
Combining (\ref{t1}) and (\ref{t2}), the temperature scales as
\begin{equation}
  T\sim (\rho_r+p_r)V_c\sim a^{-3w_r}~,
\end{equation}
so the relation between $\rho_r$ and $T$ can be written as
\begin{equation}
  \rho_r\sim T^{\frac{w_r+1}{w_r}}~.
\end{equation}
So we have the relation between the sound speed, the equation of
state, and the parameter $m$ defined in (\ref{rho}) as
\begin{equation}
  c_s^2 = w_r = \frac{1}{m-1}~.
\end{equation}

Another important issue is to determine the appropriate size of the
thermal system $L$. By determining $L$, we mean that at length
scales smaller than $L$, the fluctuation of the system can be
calculated using the thermal dynamics described above, and at scales
greater than $L$, the fluctuation is governed by the cosmological
perturbation theory. Note that a typical photon in the thermal
system has wavelength $T^{-1}$, so there is a lower bound $L\gtrsim
T^{-1}$ on $L$. Otherwise, the system is too small to be treated as
a thermal system, and the above calculation no longer holds. Also,
there should be no thermal correlation outside the acoustic horizon
$c_s H^{-1}$, so the constraint on $L$ is $T^{-1}\lesssim L\lesssim
c_s H^{-1}$. \footnote{In the literature, the length scale $T^{-1}$
is used as $L$ to calculate the thermal fluctuations by some authors
\cite{Koh:2007rx}. In our calculation of nonGaussianity, if we
choose $L=T^{-1}$, the result turns out to be much more dramatic: we
will get very large nonGaussianity for a much wider class of
inflationary models. While in our paper, we do not choose to use
$L=T^{-1}$, and only take $T^{-1}$ as an lower bound of $L$ here.}

We will argue in the discussion that from some decoupling mechanism,
the explicit value of $L$ may depend on the detailed properties of
the thermal system and the dynamics of inflation. While in the
remainder of the paper, we will hold $L$ as a parameter (sometimes
called the ``thermal horizon'') during the calculation, and discuss
the most modest limit $L=c_s H^{-1}$ when we come to final results.

\begin{figure}
\centering
\includegraphics[totalheight=1.8in]{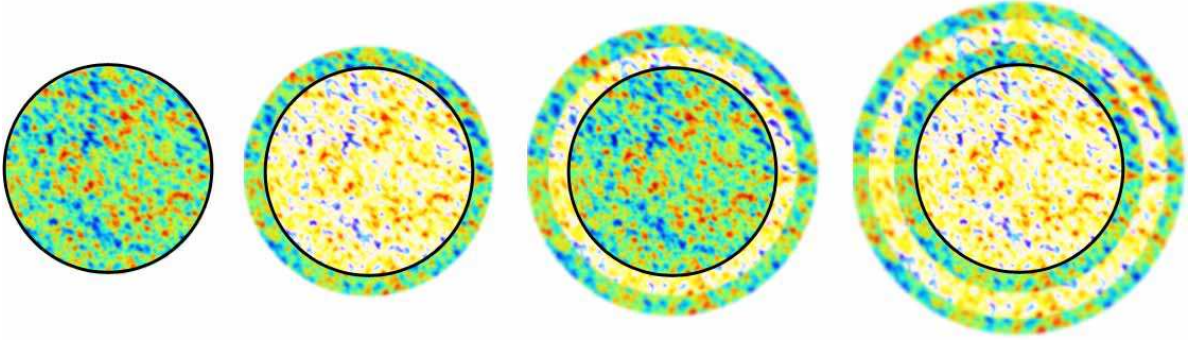}
\caption{\small{This figure illustrates how the initial condition of
perturbation is prepared by the thermal fluctuations. The black
cycle represents the thermal horizon. A fluctuation $\delta \rho$ of
the system can exit the thermal horizon $L$(shown as a shell in the
figure) during inflation. This shell outside the thermal horizon can
not return to thermal equilibrium with respect to the original
volume. It provides the initial condition of inflationary
fluctuations. Fluctuations are created shell by shell.}}
\label{fig:thermal}
\end{figure}

The fluctuation $\delta\rho$ can be thought of as an average over
all the local fluctuations of the thermal system
$\delta\rho=\frac{a^3}{V}\int d^3 x \delta\rho({\bf x})$.
Performing the fourier transformation, and linking the zero mode
of $\rm k$ to the horizon exit mode $k= a/L$, as illustrated in
Fig. \ref{fig:thermal}. Then the out-of-thermal-equilibrium
initial condition for $\delta\rho_{k}$ takes the form
\begin{equation}\label{rhok}
  \delta\rho_{k}=k^{-{3\over2}}\delta\rho~.
\end{equation}
As a check, the equation (\ref{rhok}) can also be obtained using the
window functions.

Since $L$ is smaller than the inflationary horizon, the relation
between the energy density perturbation and the scalar type metric
perturbation at the boundary of $V$ can be calculated using the
Poisson equation \footnote{Note that to be exact, the 00 component
of the linearized Einstein is $-\nabla_{\rm
ph}^2\Phi+3H\dot\Phi+3H^2\Phi=4\pi G\delta\rho$. The sound speed
$c_s$ do not enter this equation. So as long as $c_s$ is not too
large, even when $L=c_s H^{-1}$, the equation (\ref{Phik}) is a very
good approximation. }
\begin{equation}\label{Phik}
  \Phi_{{k}L} = 4\pi G \delta\rho_{k}L^2~,
\end{equation}
where $G$ is the Newton constant, and $\Phi_{{k}L}$ is the fourier
mode of the Newtonian gauge metric perturbation defined as
$ds^2=a^2\left( -(1-2\Phi)d\eta^2+(1+2\Phi)d{\bf x}^2 \right)$, and
is calculated at $k=a/L$. Note that $\Phi$ here is not the Newtonian
potential $\Phi_N$, but rather $\Phi=-\Phi_N$. We follow the WMAP
convention to use $\Phi$ as the perturbation variable. Further
discussion on the conventions can be found in the Appendix A.

We work in the Newtonian gauge only for simplicity. Since $L <
H^{-1}$, the different choice of gauge is not important inside the
thermal horizon (see, for example, \cite{Mukhanov:1990me}). After
the inflationary modes leave the thermal horizon, we use $\Phi$ to
describe the mode. It is well known that $\Phi$ can be made gauge
invariant when considering the more general gauges. So our final
result will be independent of gauge choice.

The equation (\ref{Phik}) provides a thermal initial condition of
$\Phi_{k}$. After that, the evolution of $\Phi_{k}$ is governed by
the cosmological perturbation theory. In the lowest order slow roll
approximation, $\Phi_k$ evolves as
\begin{equation}
  \Phi_k \sim \sqrt{-k\tau} H_{1/2}^{(1)}(-k\tau)~,
\end{equation}
where $\tau$ is the comoving time and $H_{1/2}^{(1)}$ is the first
kind Hankel function. It can be shown that $\Phi_{k}$ oscillates
inside the inflationary horizon with nearly constant amplitude, so
$|\Phi_{{k}}|_{k=aH}\simeq |\Phi_{{k}L}|$. After horizon crossing,
the amplitude of $\Phi_{k}$ is frozen so that the change of
$\Phi_{k}$ during a few e-folds is negligible. Using (\ref{rho2}),
(\ref{rhok}) and (\ref{Phik}), the 2-point correlation for
$\Phi_{k}$ at $k=aH$ (and also a few e-folds outside the
inflationary horizon) is expressed as
\begin{equation}\label{Phi2}
  \langle \Phi_{k}^2 \rangle = (4\pi G)^2 m\rho_r TL k^{-3}~.
\end{equation}
Note that here $k=|{\bf k}|$, and in (\ref{Phi2}), we are actually
calculating the correlation between the mode $\bf k$ and $-\bf k$.
Similarly, in the three point correlation function, the quantity we
calculate corresponds to the equilateral triangle, satisfying ${\bf
k}_1+{\bf k}_2+{\bf k}_3=0$ and $|{\bf k}_1|=|{\bf k}_2|=|{\bf
k}_3|= k$. The phase of $\Phi_k$ in the correlation functions
cancels due to the momentum conservation.

The power spectrum of $\Phi_{k}$ from the thermal origin can be
written as
\begin{equation}\label{PPhi}
  P_{\Phi}\equiv \frac{k^3}{2\pi^2} \langle\Phi_{k}^2\rangle
  = 8G^2 m\rho_r TL~.
\end{equation}

Similarly, the 3-point function of $\Phi_{k}$ at $k=aH$ is expressed
as
\begin{equation}
  \langle \Phi_{k}^3 \rangle = (4\pi G)^3 m(m+1) \rho_r T^2
  k^{-9/2}~.
\end{equation}

The nonGaussianity can be calculated for ${\zeta}_{k}$, relating to
$\Phi_{k}$ by ${\zeta}\simeq \Phi_{k}/\epsilon$ at a few e-folds
outside the inflationary horizon, where $\epsilon\equiv
-\dot{H}/H^2$.

Note that this 3-point correlation function
$\langle\zeta_k^3\rangle$ is always positive. The positivity is
transparent when we choose $L$ as the acoustic horizon $c_s
H^{-1}$. In this case, $\zeta$ freezes outside $L$ and do not have
chance to change sign. While in the case that $L< c_s H^{-1}$,
although $\zeta$ has oscillating solution inside the acoustic
horizon, but $\langle\zeta_k^3\rangle$ still can not change sign.
This is because after the initial condition is prepared, the
evolution of $\langle\zeta_k^3\rangle$ is governed by the
interaction Hamiltonian of $\zeta$ as
\begin{equation}
  \langle \zeta_k(t)^3\rangle = -i\int_{t_0}^t dt' \langle[\zeta_k(t)^3, H_{\rm
  int}(t')]\rangle~.
\end{equation}
We are assuming the slow roll inflationary scenario, and do not
employ other mechanisms to have large interaction $H_{\rm int}$. So
once the positive nonGaussian initial condition is produced, it will
keep positive until it is observed in the CMB.

Finally the nonGaussianity estimator $f_{\rm NL}$ takes the form
\begin{equation}\label{ng}
  f_{\rm NL}=\frac{5}{18} k^{-\frac{3}{2}}\frac{\langle {\zeta}_{k}^3 \rangle}
  {\langle {\zeta}_{k}^2 \rangle \langle {\zeta}_{k}^2
  \rangle}=\frac{5\epsilon (m+1)}
  {72\pi G m\rho_r L^2}~.
\end{equation}
Note that here we have assumed that the origin of perturbation is
completely thermal. A combination of the thermal and quantum
origin of the power spectrum and nonGaussianity will be discussed
at the end of this section.

From (\ref{ng}), we see that $f_{\rm NL}\propto L^{-2}$. So the
smaller $L$ is, the larger the nonGaussianity can be. Note that for
modified equation of states, if $|m|\ll 1$, the $m$ in the
denominator also enhances the nonGaussianity.

The nonGaussianity (\ref{ng}) could also be estimated without
calculating the 3-point correlation function explicitly. The idea is
to calculate the back-reaction. A fluctuation mode which crosses the
thermal horizon earlier can change the background for a later
fluctuation mode, so leads to nonGaussian correlation between these
two modes.

From the 2-point correlation function, or from the standard result
in thermodynamics that $\delta T/T\sim \sqrt{1/C_V}$, we have for
the second mode
\begin{equation}
  \delta_2\rho \sim \sqrt{mAT^{m+1}/V}~.
\end{equation}

As the first mode has crossed the thermal horizon by the time the
second mode crosses the thermal horizon, the first mode leads to a
modification of the background of the thermal system for the second
mode. This modification of the background represents the correlation
of the two modes, so is the nonGaussian contribution. At the thermal
horizon, this nonGaussian contribution takes the form
\begin{equation}
  \delta_1(\delta_2\rho)\sim \frac{m+1}{2}\sqrt{mAT^{m-1}/V}\delta_1
  T\sim \frac{(m+1)T}{2V}~.
\end{equation}

As discussed earlier, when the modes reaches the inflationary
horizon $k=aH$, we have
\begin{equation}
  {\zeta}-{\zeta}_g\sim 4\pi G \delta \rho L^2/ \epsilon~.
\end{equation}
So finally the nonGaussianity $f_{\rm NL}$ is estimated as
\begin{equation}
   f_{\rm NL}\sim \frac{5({\zeta}-{\zeta}_g)}{3{\zeta}_g^2}\sim \frac{5\epsilon (m+1)}
  {24\pi G m\rho_r L^{2}}~,
\end{equation}
This differs from (\ref{ng}) only by a factor of 3, and can be
considered as in good agreement for a rough estimate. This
back-reaction estimate provides a check for the 3-point function
calculated above, and also explains why the nonGaussianity can be so
large: the thermal horizon is smaller than the inflationary horizon,
so a back-reaction calculated at the thermal horizon is larger than
the one calculated at the inflationary horizon. This large
back-reaction leads to a large nonGaussianity. Note that although
this estimate can not give the precise shape of $f_{\rm NL}$, the
limit we take is similar to the squeezed limit, which leads to a
local shape nonGaussianity.

Generally, there can also be perturbations from the vacuum
fluctuations of the coherent rolling inflaton field. Let us denote
this perturbation by $\Phi_{{k}}^{\rm vac}$. Since the vacuum
fluctuation and the thermal fluctuation are of the different origin,
they do not have correlations between each other. So in the 2-point
and 3-point functions, the cross terms such as $\langle
\Phi_{k}\Phi_{{\bf k}}^{\rm vac}\rangle$ vanishes. So for the total
power spectrum and the nonGaussianity,
\begin{equation}\label{sumrule}
  P_{\Phi}^{\rm tot}= P_{\Phi}^{\rm vac}+P_\Phi~, ~~~ f_{\rm NL}^{\rm
  tot}=\frac{5}{18} k^{-\frac{3}{2}}\frac{\langle {\zeta}_{{k}}^{{\rm vac}~3} \rangle+\langle {\zeta}_{k}^3 \rangle}
  {\left\{\langle {\zeta}_{{k}}^{{\rm vac}~2} \rangle +\langle {\zeta}_{k}^2
  \rangle\right\}^2}~,
\end{equation}
where ${\zeta}_{{k}}^{\rm vac}$ and $P_{\Phi}^{\rm vac}$ are the
comoving curvature perturbation and the power spectrum calculated
from the vacuum fluctuation of the inflaton field. When
$|\Phi_{{k}}^{\rm vac}| \ll |\Phi_{k}|$, (\ref{sumrule}) returns to
(\ref{PPhi}) and (\ref{ng}), and when $|\Phi_{{k}}^{\rm vac}| \gg
|\Phi_{k}|$, (\ref{sumrule}) returns to the power spectrum and the
nonGaussianity with zero temperature.

\section{Examples of inflation models with large thermal nonGaussianity}

In the previous section, we have given the general formalism to
calculate the thermal perturbations and nonGaussianity. In this
section, we apply the formalism to the chain inflation and the warm
inflation.

Although the radiation energy density can be inflated away very
easily, as discussed in the introduction, there are mechanisms to
continuously produce radiation so that the radiation energy density
keeps nearly constant. Such mechanisms include the interaction of
the radiation with the inflaton, and the bubble collision during the
chain inflation, which we shall show explicitly.

In the chain inflation model, the vacua tunnel rapidly and the time
evolution of the vacuum energy density can be approximated by
\begin{equation}\label{rhot}
  \rho_0(t)=\rho_0(0)- \alpha t~,
\end{equation}
where $\alpha$ denotes the averaged decay rate of the vacuum
energy. ($\alpha\equiv \frac{\sigma}{\tau}$ in the notation of
\cite{Huang:2007ek}). Suppose that the decreasing energy converts
completely into the radiations through bubble collision. Taking
into consideration of the red shift of the radiation during
inflation, the radiation energy density satisfies
\begin{equation}\label{de}
  d\rho_r(t)=\alpha dt -3H(1+w_r)\rho_r dt~.
\end{equation}

By taking the stationary limit $t\gg [(1+w_r)H]^{-1}$, we have
\begin{equation}\label{rhor}
  \rho_r=\frac{\alpha}{3(1+w_r)H}=
  \frac{2\epsilon\rho}{3(1+w_r)}~,~~~\epsilon\equiv -\frac{\dot
  H}{H^2}=\frac{4\pi G}{3}\frac{\alpha}{H^3}~.
\end{equation}
The relation (\ref{rhor}) could also be obtained from assuming that
the radiation density is produced within one Hubble time, then from
(\ref{rhot}) and taking $t\sim H^{-1}$, we get directly that
$\rho_r$ is of the order $\alpha/H$.

Note that $\rho_r$ is a slow roll quantity during inflation. As
$\rho_r\sim T^m$, when $m$ is not too small, $T$ also changes very
slowly during inflation. This verifies the assumption that the
radiation density and the temperature during inflation are almost
constants.

Several mechanisms have been proposed to calculate the fluctuations
in the chain inflation model. In \cite{Feldstein:2006hm}, the
authors showed that the perturbations can come from different pathes
along which the meta-stable vacuum tunnel. In \cite{Huang:2007ek},
the density perturbation is calculated by applying the standard
formulism to the effective scalar field characterizing the
tunnelling effect. But up to now, the exact mechanism how the
perturbation in chain inflation is generated is still not clear.

In this section, we propose a new mechanism to produce the density
perturbation of the chain inflation. We claim that the perturbation
can have a thermal origin. It is because after the bubble collision,
the energy contained in the bubble wall becomes the radiation. As to
be shown later in this section, the radiation density is of order
$\rho_r\sim\epsilon \rho$, whose thermal fluctuation can exit the
horizon and produce a scale invariant power spectrum. So in a
realistic calculation,  the fluctuations discussed in
\cite{Feldstein:2006hm}, \cite{Huang:2007ek} and the thermal
fluctuation should be taken into account at the same time.

Using (\ref{PPhi}), the power spectrum takes the form
\begin{equation}
  P_\Phi= \frac{16G^2 m\epsilon \rho TL}{3(1+w_r)}~,~~~P_\zeta=\frac{16G^2 m \rho TL}{3(1+w_r)\epsilon}~.
\end{equation}
Assuming $m$ and $w_r$ are exactly constants, the spectral index
takes the form
\begin{equation}
  n_s-1\equiv \left.\frac{d\ln P_{\zeta}}{d\ln
  k}\right|_{k=aH}=-5\epsilon +\frac{1}{TL}\frac{d(TL)}{Hdt}
\end{equation}

If $L$ saturates its lower bound $L\sim T^{-1}$, and considering the
usual type of radiation $m=4$, $w_r=1/3$, then we recover the power
spectrum and the spectral index in the simplified chain inflation
model \cite{Huang:2007ek}.

If the thermal perturbation is dominate over other perturbation
sources during chain inflation, then the nonGaussianity can be read
off from (\ref{ng}) that
\begin{equation}
  f_{\rm NL}= \frac{5(m+1)(1+w_r)}{18m(LH)^2}~.
\end{equation}
Note that for the usual type of radiation $m=4$, $w_r=1/3$, $L$
should satisfy $L\leq (3H)^{-1}$, so $f_{\rm NL}$ is always larger
than ${\cal O}(1)$. This is very different from the ordinary
inflation model in which $f_{\rm NL}\sim {\cal O}(\epsilon)$.

To go one step further, let us consider the modified radiation
$m\neq 4$. And make a modest estimate that $L=c_s H^{-1}$. In this
case,
\begin{equation}
  f_{\rm NL}=\frac{5(m+1)(1+w_r)}{18m c_s^2}=\frac{5(m+1)}{18}
\end{equation}
As a phenological model, if $m$ is large, one can have large $f_{\rm
NL}$.

If the thermal perturbation is not the dominate source in the power
spectrum, then we need to compare the thermal and other
contributions to estimate the nonGaussianity. Let the power spectrum
from the other origin be $P_{\Phi}^{\rm vac}$, then using (\ref{ng})
and (\ref{sumrule}), where we have neglected the nonGaussianity
produced by the sources other than thermal, we can express $f_{\rm
NL}$ as
\begin{equation}\label{ng1}
f_{\rm NL}= \frac{5(m+1)(1+w_r)}{18m(LH)^2(1+\frac{P_{\Phi}^{\rm
vac}}{P_\Phi})^2}~.
\end{equation}
So it is clear that although the nonGaussianity is suppressed by
the ratio of the spectrums, there is no longer the $\epsilon$
suppression. Moreover, if $L \ll H^{-1}$, then the nonGaussianity
can be enhanced by a great amount.

In the case of warm inflation, the radiation is continuously
produced during slow roll inflation. This process can be modeled by
adding a interacting term between inflaton and radiation component
in the Lagrangian. In the slow roll regime, the equation of motion
for inflaton $\varphi$ is
\begin{equation}
  3H\dot\varphi + \Gamma_\varphi \dot\varphi + \partial_{\varphi} V(\varphi) = 0~,
\end{equation}
where $\Gamma_\varphi$ is the decay rate for the inflaton to
radiation process. We assume $\Gamma_\varphi$ is a constant (or at
least a slow roll quantity) here. The equation for the radiation
energy density takes the form
\begin{equation}
  \dot\rho_r+3H(1+w_r)\rho_r=\Gamma_\varphi \dot\varphi^2~.
\end{equation}
A solution for these equations is given by
\begin{equation}\label{warmsolution}
  \dot\rho_r \simeq 0~,~~~ \rho_r\simeq
  \frac{\Gamma_\varphi}{3H(1+w_r)}\dot\varphi^2~.
\end{equation}
And in \cite{Berera:1995wh}, it is shown that this solution is an
attractor solution, and independent of the initial conditions for
the thermal component.

When $\rho_r \lesssim\rho_0$, the universe accelerates. To give a
nearly scale invariant spectrum, we require the universe undergoes a
quasi-dS expansion, so the bound for the radiation energy should be
$\rho_r \lesssim \epsilon \rho_0$. This corresponds to the case that
$\Gamma_\varphi \lesssim H$. When this bound saturates, the power
spectrum and the nonGaussianity coincides with the chain inflation
case.

It can be checked that when the constraint $\rho_r \lesssim \epsilon
\rho_0$ is satisfied, the e-folding number and the slow roll
condition is qualitatively the same as the $\rho_r=0$ case, and the
inflaton vacuum is not thermalized because the interaction rate
$\Gamma_\varphi<H$. So the inflationary background and the amplitude
of inflaton fluctuation do not change. In \cite{Berera:1995wh}, the
authors also considered the case $\rho_r \gtrsim \epsilon \rho_0$,
but we will not investigate this case in detail here, and assume
$\rho_r \lesssim \epsilon \rho_0$ from now on.

Similarly to the chain inflation case, the thermal power spectrum of
warm inflation takes the form
\begin{equation}
  P_\Phi= 8G^2 m \rho_r TL~,~~~ P_{\zeta} = 8G^2 m\rho_r
  TL/\epsilon^2~.
\end{equation}

And the nonGaussianity of the warm inflation is
\begin{equation}\label{ngt}
  f_{\rm NL}=
  \frac{5(m+1)\epsilon\rho}{27m\rho_r (LH)^2 (1+\frac{P_\Phi^{\rm vac}}{P_\Phi})^2}~,
\end{equation}
which behaves like the chain inflation model with sub-dominate
thermal fluctuation. Note that $\rho_r \lesssim \epsilon \rho_0$, so
there is no ${\cal O}(\epsilon)$ suppression. The enhancement due to
$(LH)^{-2}$ still exists.

To see how $f_{\rm NL}$ behaves at small $\rho_r$ limit, taking into
consideration the observational constraint of the total power
spectrum $P_{\zeta}^{\rm tot}\simeq 2.5 \times 10^{-9}$, $f_{\rm
NL}$ can be written as
\begin{equation}
  f_{\rm NL}=\frac{40 m (m+1) G^3 \rho_r T^2}{9\pi P_\zeta^{{\rm
  tot}2}\epsilon^3}~.
\end{equation}
So when taking the $\rho_r\rightarrow 0$ limit, we get $f_{\rm
NL}\rightarrow 0$. It is reasonable because the $f_{\rm NL}$ we are
considering comes from thermal origin.

It seems surprising at first sight that in the limit that
$\epsilon$ is very small, the radiation component bounded by
$\rho_r\lesssim \epsilon \rho$ is tiny, but the thermal
nonGaussianity can still be quite large. The reason for this is
that given the amplitude of the observed CMB power spectrum, when
$\epsilon$ is very small, the primordial density fluctuation
needed to generate the CMB power spectrum also becomes small. So a
tiny part of radiation becomes comparable with the inflaton vacuum
fluctuation in the function of generating the fluctuations.

\section{Conclusion and discussion}

As a conclusion, in this paper, we have calculated the
nonGaussianity from thermal effects during inflation. We
calculated the 2-point and 3-point thermal correlation functions,
and using these correlation functions to calculate the scalar
power spectrum $P_{\Phi}$ and the nonGaussianity estimator $f_{\rm
NL}$. We also used an independent method to check the value of
$f_{\rm NL}$.

We have applied our treatments in the chain inflation. We find that
the density perturbation in the chain inflation may come from the
thermal fluctuations. This provides a candidate for the origin of
power spectrum of the chain inflation.

We also calculated the nonGaussianity of the chain inflation model.
We found if the thermal perturbation is the main source of chain
inflation, then the nonGaussianity $f_{\rm NL}$ of chain inflation
is greater than ${\cal O}(1)$. Taking into consideration the
modified sound speed, the nonGaussianity can become much larger.

If the thermal perturbation is sub-dominate during chain inflation,
then $f_{\rm NL}$ is suppressed by $P_{\Phi}^2/(P_{\Phi}+P^{\rm
vac}_{\Phi})^2$. But still, there is no ${\cal O}(\epsilon)$
suppression, and the term $(LH)^{-2}$ can provide a large
nonGaussianity.

As another application, we studied the nonGaussianity in the warm
inflation model. The result of the warm inflation model is similar
to the case of chain inflation with sub-dominate thermal component.

We only studied the $\rho_r \lesssim \epsilon\rho_0$ case in the
warm inflation scenario. It is shown that large nonGaussianity can
already show up in this case. We have not considered in this paper
the complementary case $\rho_r > \epsilon\rho_0$. In this latter
case, the thermalization of the inflaton vacuum dominates the
power spectrum. The thermal part of the nonGaussianity $f_{\rm
NL}^{\rm thermal}$ is suppressed in this case. But the thermal
nonGaussianity of the inflaton field should be taken into
consideration.

In the warm inflation case, it is clear that the vacuum fluctuation
and the thermal fluctuation are two different sources of inflation
fluctuations. So it may lead to large isocurvature perturbation.
This isocurvature perturbation issue is not discussed in detail in
this paper.

In this paper, we calculated the equilateral shape nonGaussianity.
And in the back reaction estimate, the shape is something like local
shape. Since the nonGaussianity from thermal fluctuations can be
large, and is very hopeful to be observed in the near future, it is
also important to calculate the more general correlation functions
with arbitrary ${\bf k}$, and obtain the shapes of the
nonGaussianity.

Another important issue is to determine the parameter $L$ given a
inflation model, which requires more details on the dynamics of
the system. In this paper, we mainly discussed the upper bound of
$L$, which is governed by the sound speed $c_s$ of the radiation
component. But we note that $L$ may be much smaller than the
acoustic horizon. One possible mechanism generating smaller $L$ is
the decoupling of the fourier mode of the thermal fluctuations.
When the universe expands, the interaction rate $\Gamma$ for the
thermal fluctuation fourier mode may decrease, so it decouples
before reaching the acoustic horizon. We wish we will address this
issue in the near future.

The generalization of our calculation to other inflation models with
radiation is straightforward. For example, our calculation can also
be applied to the thermal inflation model \cite{Lyth:1995hj}, or the
thermal version of the noncommutative inflation model
\cite{Koh:2007rx}.

The similar analysis can also be performed in the string gas model
\cite{Brandenberger:1988aj}, where the power spectrum also has a
thermal origin. The calculation of the nonGaussianity of the
string gas model will be represented in a separate publication
\cite{preparation}.

\section*{Acknowledgments}

We would like to thank Robert Brandenberger, Eugene A. Lim, Chuan
Liu, Bo-Qiang Ma, Henry Tye, Zhi-Xin Qian, J. Soda, and especially
Yi-Fu Cai, Xingang Chen, Qing-Guo Huang and Eiichiro Komatsu for
discussion. We are grateful to KITPC for its wonderful programme on
String and Cosmology, where much of this work was done and
discussed. BC would like to thank OCU for its hospitality, where
this project was finished. The work was partially supported by NSFC
Grant No. 10535060, 10775002.

\section*{Appendix A: Clarification on Conventions}

In this appendix, we clarify the convention we use. To write an
appendix to clarify the convention is necessary, because a confusion
in the convention (especially the sign) can lead to an extra minus
sign in $f_{\rm NL}$, and lead to completely opposite predictions.
This is very different from the calculation of the power spectrum,
where a confusion of the sign convention usually leads to the same
result.

In this paper, we use the WMAP convention. In this convention, the
metric perturbation $\Phi$ can be written as (in the Newtonian
gauge)
\begin{equation}
  ds^2 = a^2 \left(-(1-2\Phi)) d\eta^2 +(1+2\Phi) d {\bf x}^2
  \right)~.
\end{equation}
So $\Phi$ is not the Newtonian potential $\Phi_N$, but rather
$\Phi=-\Phi_N$. This is the same as the convention used in
\cite{Komatsu:2002db}, which is also the same as the convention
used in the WMAP group. One can refer to \cite{Komatsu:2002db} to
find the complete definitions. This is of the different sign from
the convention $\Phi$ used in \cite{Mukhanov:1990me}. (In
\cite{Mukhanov:1990me}, they use $\phi$ to denote the Newtonian
potential, and the gauge invariant quantity $\Phi=\phi$ in the
Newtonian gauge.)

For the quantity $\zeta$, there are also different conventions in
the literature. In this paper, following the convention of
\cite{Komatsu:2002db}, $\zeta$ can be written as
\begin{equation}
  \zeta = \Phi - \frac{H}{\partial_t \varphi} \delta\varphi~,
\end{equation}
where $\varphi$ and $\delta\varphi$ are the background value and the
perturbations for the inflaton field respectively. And in
\cite{Mukhanov:1990me}, the $\zeta$ parameter they use is of the
different sign.

One simple way to check the sign is to relate it to the quantities
which have clear physical meaning. There are at least two such
quantities: the energy density $\delta \rho$ and the CMB
temperature fluctuation $\Delta T/T$. In our convention, the
Poisson equation takes the form
\begin{equation}
  -\frac{\nabla^2}{a^2}\Phi = 4\pi G \delta \rho~.
\end{equation}
And the CMB temperature fluctuation can be written as
\begin{equation}
  \frac{\Delta T}{T}=-\frac{1}{3}\Phi=-\frac{1}{5}\zeta~.
\end{equation}

Although not related to this paper, we also would like to remind
the reader two more differences in the conventions, which may be
used in the calculation of $f_{\rm NL}$. One is that in \cite{NG},
Maldacena uses the same convention of $\zeta$ as that of the WMAP
group, but the equation in the footnote 16,
$\zeta=-\frac{5}{3}\Phi$, does not follow the WMAP convention. So
the $f_{\rm NL}$ defined in \cite{NG} is of the different sign
from the WMAP convention. The other is that for the so called
``comoving curvature perturbation'' $\cal R$. The $\cal R$ used in
\cite{Komatsu:2002db} (in the comoving gauge) is of the different
sign from the $\cal R$ used in \cite{Riotto:2002yw} outside the
horizon.

\end{document}